# A new method for analyzing the time evolution of quantum mechanical systems

Chyi-Lung Lin[1] and Tsin-Fu Jiang[2]
[1]Department of Physics, Soochow University,
Taipei, Taiwan, R.O.C.
[2] Institute of Physics, National Chiao Tung University,
Hsinchu, 30010, Taiwan

## ABSTRACT

We show a new method for analyzing the time evolution of the Schrödinger wave function $\Psi(x, t)$. We propose the decomposition of the Hamiltonian as: $H(t) = \widetilde{H}(t) + H_c(t)$, where $\widetilde{H}(t)$ is the operator which does not change the state and therefore $\Psi(x, t)$ is its eigenfunction, and $H_c(t)$ is the operator that changes the state. With this decomposition, the time evolution of a wave function can be understood more clearly via the operator $H_c(t)$. We illustrate this method by exactly solving the system of driven harmonic oscillator. We show that nonspreading wave packets exist in this system in addition to historically known paradigms. This method can be applied to analyze the time evolution of general Hamiltonian systems as well.

## 1. Introduction

It is the Hamiltonian that governs the change in time of the wave function $\Psi(x, t)$. This is described in the Schrödinger equation

$$i\hbar\, \partial_t \Psi(x, t) = H(t)\, \Psi(x, t) \qquad (1)$$

We discuss the general case that the Hamiltonian may be time dependent. The action of $H(t)$ on $\Psi(x, t)$ in general is complicated. The simplest case is when $\Psi(x, t)$ is the instantaneous eigenfunction of $H(t)$; then $H(t)\, \Psi(x, t) = E(t)\, \Psi(x, t)$, where $E(t)$ is the instantaneous eigenvalue. Thus, when a state evolves in time, it would be helpful at each instant to know those operators which do not change the time-evolved state. The situation that an operator does not change a state corresponds to an eigenvalue equation.

To find eigenvalue equations of $\Psi(x, t)$, we consider the wave evolving from an initial wave $\Psi(x, 0)$. We have the following relation

$$\Psi(x, t) = U(t, 0)\, \Psi(x, 0) \qquad (2)$$

where $U(t,0)$ is the time evolution operator. $U(t,0)$ is unitary with Hermitian Hamiltonian $H(t)$. By means of time evolution operator, we can carry out a unitary transformation of states, operators and eigenvalue equations.

The operators chosen to define the eigenvalue equations are case dependent. We denote such an operator by $\tilde{H}$ which may represent a position operator, a momentum operator or an energy operator, etc. Thus, by the time evolution operator, $\Psi(x, 0)$ is transformed to $\Psi(x, t)$, and an operator $\tilde{H}$ at time $t = 0$, denoted by $\tilde{H}(0)$, is transformed to $\tilde{H}(t)$ which is defined by

$$\tilde{H}(t) = U(t, 0)\, \tilde{H}(0)\, U^{-1}(t, 0) \qquad (3)$$

Suppose at time $t = 0$, we have an initial wave function $\Psi(x, 0)$ and its corresponding eigenvalue equation:

$$\tilde{H}(0)\, \Psi(x, 0) = \tilde{E}(0)\, \Psi(x, 0) \qquad (4)$$

where $\tilde{E}(0)$ is the corresponding eigenvalue. This eigenvalue equation at time t = 0 can be transformed to an eigenvalue equation at time t by the unitary transformation. This is done by operating both sides of (4) by U (t, 0), we then obtain

$$\tilde{H}(t)\,\Psi(x,t) = \tilde{E}(t)\,\Psi(x,t) \qquad (5)$$

where

$$\tilde{E}(t) = \tilde{E}(0) \equiv \tilde{E} \qquad (6)$$

Eq. (5) is the eigenvalue equation of $\Psi(x,t)$. We can determine $\tilde{H}(t)$ from (3) and calculate $\Psi(x,t)$ from Eq. (5). It may occur that solving Eq. (5) is easier than solving Eq. (1)

In all, what we are managing is the establishment of two eigenvalue equations, one is at time t = 0, and the other is at time t. We can use this time development of eigenvalue equation to explore the time development of wave function.

We call $\tilde{H}(t)$ the *state-preserving operator*, as $\tilde{H}(t)$ does not change the state. Let $H_c(t) = H(t) - \tilde{H}(t)$. The Hamiltonian is then decomposed into

$$H(t) = \tilde{H}(t) + H_c(t) \qquad (7)$$

Eq. (7) means that we divide the Hamiltonian H(t) into two parts, where $\tilde{H}(t)$ is the operator which does not change the state, and $H_c(t)$ is the operator which actually changes the state. We call $H_c(t)$ the *state-changing operator*. The suffix c is referred to "change". This decomposition of the Hamiltonian is different from that of the interaction picture in which the Hamiltonian is split into: $H(t) = H_0 + H_{int}(t)$. Our decomposition is dynamical, as the time development of $\tilde{H}(t)$ follows the time development of the wave function $\Psi(x,t)$.

From (7), Schrödinger equation can then be written in a simpler form as below:

$$i\hbar\,\partial_t\,\Psi(x,t) = H(t)\,\Psi(x,t)$$
$$= [\,\tilde{E}(t) + H_c(t)\,]\,\Psi(x,t) \qquad (8)$$

The time evolution of a wave function can be understood more clearly via the Hamiltonian $H_c(t)$. Eq. (8) is the formula we use to solve $\Psi(x,t)$. It may occur that solving Eq. (8) is easier than solving Eq. (1) or Eq. (5).

We found that the decomposition of Hamiltonian is particularly interesting when it is applied to analyzing the time evolution of nonspreading wave packets (NSWPs) [4]. In 1926, Schrödinger constructed the first NSWP with the profile of the ground state of simple harmonic oscillator (SHO) [1]. This example shows that NSWPs exist in a system with Hamiltonian $H(t) = \frac{p^2}{2m} + \frac{1}{2}m\omega^2 x^2$. In 1954 Senitzky generalized Schrödinger's result, constructing NSWPs with the profiles of high energy eigenstates of SHO [2]. Other type of NSWPs was found in 1979 by Berry and Balazs [3]. They showed that NSWPs exist in a system with Hamiltonian $H(t) = \frac{p^2}{2m}$, that is the free space, and also in a system with Hamiltonian $H(t) = \frac{p^2}{2m} - F[t]\, x$. Comparing these two types of NSWPs, we may suspect that NSWPs should also exist in a system with $H(t) = \frac{p^2}{2m} + \frac{1}{2}m\omega^2 x^2 - F[t]\, x$. This is the driven or forced harmonic oscillator.

In Sec. 2, we apply this decomposition method to exactly solving $\Psi(x,t)$ of the driven harmonic oscillator. Our result can be compared to known solution [5]. However, we obtain the exact solution for broader initial conditions and particularly we show the existence of NSWPs in this system. In Sec. 3, we perform numerical simulations to justify the theoretical results. In Sec. 4, we draw a brief conclusion.

2. **Driven simple harmonic oscillator**

We discuss the time evolution of the driven harmonic oscillator. The Hamitonian is defined as below:

$$H(t) = \frac{p^2}{2m} + \frac{1}{2}m\omega^2 x^2 - F(t)\, x \qquad (9)$$

where F(t) is an arbitrary function of time. The eigenvalue equation of the simple harmonic oscillator is

$$\left(\frac{p^2}{2m} + \frac{1}{2}m\omega^2 x^2\right)\Phi_n(x) = E_n\Phi_n(x). \qquad (10)$$

where $E_n = \left(n + \frac{1}{2}\right)\hbar\omega$, $n = 0, 1, 2, \ldots$. We take (10) as the initial eigenvalue equation, then

$$\Psi(x, 0) = \Phi_n(x) \quad (11)$$

$$\tilde{H}(0) = \left(\frac{p^2}{2m} + \frac{1}{2}m\omega^2 x^2\right) \quad (12)$$

We introduce the two quantities as follows

$$x_t = U(t, 0)\, x\, U^{-1}(t, 0) \quad (13)$$

$$p_t = U(t, 0)\, p\, U^{-1}(t, 0) \quad (14)$$

where

$$U(t, 0) = \prod_{i=0}^{N} \mathrm{Exp}\left[\frac{-i}{\hbar} H(t_i)\, \Delta t\right] \quad (15)$$

is the time evolution operator. In Eq. (15), N is a large number, $\Delta t = t/N$, $t_i = i\,\Delta t$. Then $\tilde{H}(t)$ and the eigenvalue equation of $\Psi(x, t)$ obtained from unitary transformation are as followings

$$\tilde{H}(t) = \left(\frac{p_t^2}{2m} + \frac{1}{2}m\omega^2 x_t^2\right) \quad (16)$$

$$\tilde{H}(t)\,\Psi(x, t) = E_n \Psi(x, t) \quad (17)$$

From (13)-(15), we can calculate $x_t$ and $p_t$ and obtain the following results

$$x_t = \cos[\omega t]\, x - \frac{\sin[\omega t]}{m\omega} p + \frac{1}{m\omega} fs[t] \quad (18)$$

$$p_t = m\omega \sin[\omega t]\, x + \cos[\omega t]\, p - fc[t] \quad (19)$$

where

$$fs[t] = \int_0^t F(\tau)\sin[\omega \tau]\, d\tau \quad (20)$$

$$fc[t] = \int_0^t F(\tau)\cos[\omega \tau]\, d\tau \quad (21)$$

Then we can calculate $\tilde{H}(t)$ from (16). It is more convenient to introduce the quantity d(t) defined below

$$d(t) = \frac{1}{m\omega}\int_0^t F(\tau) \sin[\omega(t-\tau)]\, d\tau \qquad (22)$$

From Eq. (22), we have $d(0) = 0, \dot{d}(0) = 0$. The meaning of d(t) will be discussed below. Substituting (18)-(19) into (16), we obtain

$$\tilde{H}(t) = \frac{p^2}{2m} + \frac{1}{2}m\omega^2 x^2 - \dot{d}(t)\, p + [m\ddot{d}(t) - F(t)]\, x +$$

$$\frac{1}{2m}(f_c[t]^2 + f_s[t]^2) \qquad (23)$$

The Hamitonian can then be decomposed into the following

$$H(t) = \tilde{H}(t) + H_c(t) \qquad (24)$$

$$H_c(t) = \dot{d}(t)\, p - m\ddot{d}(t)\, x - \frac{1}{2m}(f_c[t]^2 + f_s[t]^2) \qquad (25)$$

The form of the operator $H_c(t)$ in (25) is an indication of NSWPs [4]. From (24)-(25), Schrödinger equation can then be written in the following form

$$i\hbar\, \partial_t \Psi(x,t) = [\, E_n + H_c(t)\, ]\, \Psi(x,t) \qquad (26)$$

Eq. (26) is an equation linear in x and p, we can easily solve it and obtain the result as follows

$$\Psi(x,t) = \text{Exp}\left[\frac{-i}{\hbar}\varphi(x,t)\right] \text{Exp}\left[\frac{i}{\hbar} m\dot{d}(t)\, x\right] \text{Exp}\left[\frac{-i}{\hbar} d(t)\, p\right] \Psi(x,0)$$

$$(27)$$

where

$$\varphi(x,t) = \int_0^t \left(E_n - \frac{1}{2m}(f_c[s]^2 + f_s[t]^2) - m\dot{d}(t)^2\right) dt \qquad (28)$$

The operator $\text{Exp}\left[\frac{-i}{\hbar} d(t) p\right]$ in (27) is a spatial shift operator; therefore,

the shape of the wave function $\Psi(x,t)$ is the same as that of $\Psi(x,0)$. We conclude the result as the following:

$$\Psi(x,0) = \Phi_n(x)$$
$$\Psi(x,t) = \exp\left[\frac{-i}{\hbar}\varphi(x,t)\right] \exp\left[\frac{i}{\hbar} m \dot{d}(t) x\right] \Phi_n(x-d(t)) \qquad (29)$$

It shows that $\Psi(x,t)$ is an NSWP, and its motion is described by the trajectory: $x = d(t)$. It is interesting to note that $d(t)$ in fact is the solution of the equation:

$$\ddot{d}(t) + \omega^2 d(t) = F(t)/m \qquad (30)$$

Eq. (30) is just the classical equation obtained from the Hamiltonian defined in (9). This shows that this NSWP behaves just like a classical particle whose motion obeys the classical dynamics of the Hamiltonian H(t). We thus get the illustrative result that the driven simple harmonic system does allow the existence of NSWPs, no matter how complicated the driving force F[t] is. It seems that this result has not been discussed before to our knowledge.

We can understand why NSWPs exist in this driven system. From (26), we see that the time evolution is in fact governed by $H_c(t)$. From (25), we see that $H_c(t)$ is linear in x and p. This means that the time evolution operator is essentially a spatial-shift operator. Hence the wave packet evolves in time without distortion. We may say that $H_c(t)$ is the effective Hamitonian governing the time evolution of wave packets.

Taking n = 0 in (29), then $\Psi(x,t)$ is proportional to $\Phi_0(x-d(t))$, which represents a displaced ground state, and is known as a coherent state. This leads to the well-known result that the probability distribution among states is a Poisson distribution.

## 3. Numerical simulations

To justify the developed Hamiltonian decomposition method, we consider a ground state electron moving in harmonic potential and being excited by a Sine-squared laser pulse with carrier frequency $\Omega$. That is

$$F(t) = F_m \sin\left(\frac{\pi t}{T}\right)^2 \sin(\Omega t) A, \quad \text{for } 0 \leq t \leq T \qquad (31)$$
$$= 0, \quad \text{otherwise} \qquad (32)$$

The corresponding d(t) in (22) can be derived explicitly. For nonresonant excitation $(\Omega \neq \omega)$ and $0 \leq t \leq T$, we have (with $\Lambda = \frac{2\pi}{T}$)

$$d(t) = \frac{F_m}{2 m \omega (\omega^2 - \Omega^2)} (\omega \sin[\Omega t] - \Omega \sin[\omega t]) - \frac{F_m}{4 m \omega}$$

$$\left\{ \frac{(\Lambda - \Omega) \sin[\omega t]}{\omega^2 - (\Lambda - \Omega)^2} - \frac{(\Lambda + \Omega) \sin[\omega t]}{\omega^2 - (\Lambda + \Omega)^2} - \frac{\omega \sin[(\Lambda - \Omega) t]}{\omega^2 - (\Lambda - \Omega)^2} + \frac{\omega \sin[(\Lambda + \Omega) t]}{\omega^2 - (\Lambda + \Omega)^2} \right\} \quad (33)$$

For resonant excitation $(\omega = \Omega)$,

$$d(t) =$$

$$\frac{F_m}{4 m \omega} \left\{ \frac{\sin[\omega t]}{\omega} - t \cos[\omega t] - \frac{2\omega \sin[\omega t]}{4\omega^2 - \Lambda^2} - \frac{\omega \sin[(\omega - \Lambda) t]}{\Lambda (2\omega - \Lambda)} \frac{\omega \sin[(\omega + \Lambda) t]}{\Lambda (2\omega + \Lambda)} \right\}$$

$$(34)$$

We discretize the system by the Fourier-Grid-Hamiltonian method [6]. Jiang [7] has applied the method to study the high-frequency stabilization of an excited Morse oscillator under intense fields. We integrate the time-dependent Schrödinger equation by the variable-step Adams method [8]. During the excitation, we calculate the energy expectation of the oscillator by

$$\langle H(t) \rangle = \langle \frac{p^2}{2m} \rangle + \langle \frac{1}{2} m \omega^2 x(t)^2 \rangle - F(t) \langle x(t) \rangle \quad (35)$$

And the acceleration function through the Ehrenfest's theorem :

$$m \langle \ddot{x}(t) \rangle = \langle -\frac{\partial H}{\partial x} \rangle = F(t) - m \omega^2 \langle x(t) \rangle \quad (36)$$

Also we calculated the uncertainty $\Delta x$, $\Delta p$ and the product $\Delta x \Delta p$ during the excitation duration. In Fig.1a, we present the results of typical nonresonant excitation,

$$\Omega = \frac{\omega}{2}, F_m = 1, T = 10 \text{ cycle (cycle} = \frac{2\pi}{\omega}) \quad (37)$$

The classical analytical form d(t) (red line), the peak of wave packet (blue dots) together with the laser pulse F(t) (black line) are shown. We can see the peak of wave packet agrees with the classical d(t). And the motion is in phase with the driving pulse F(t). Fig. 1b shows the time behaviors of uncertainty $\Delta x, \Delta p$ and the product $\Delta x \Delta p$. The minimum uncertainty (coherent state) was shown through the excitation. In Fig. 1c, we depict the $\langle H(t) \rangle$ and the acceleration with respect to time. The total energy oscillates with time follows the pulse. Although the energy oscillates, but the wave packet keeps nonspreading as can be seen in Fig.1b.

Fig.1a

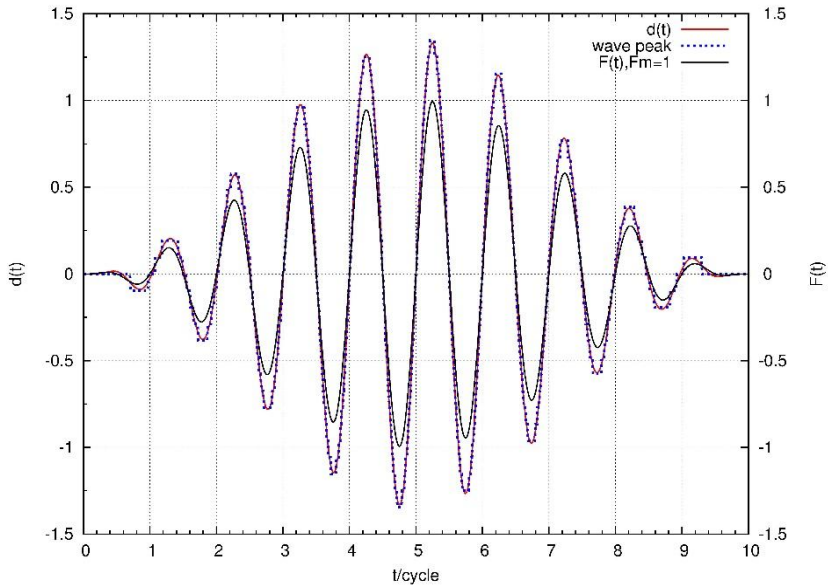

Fig. 1b

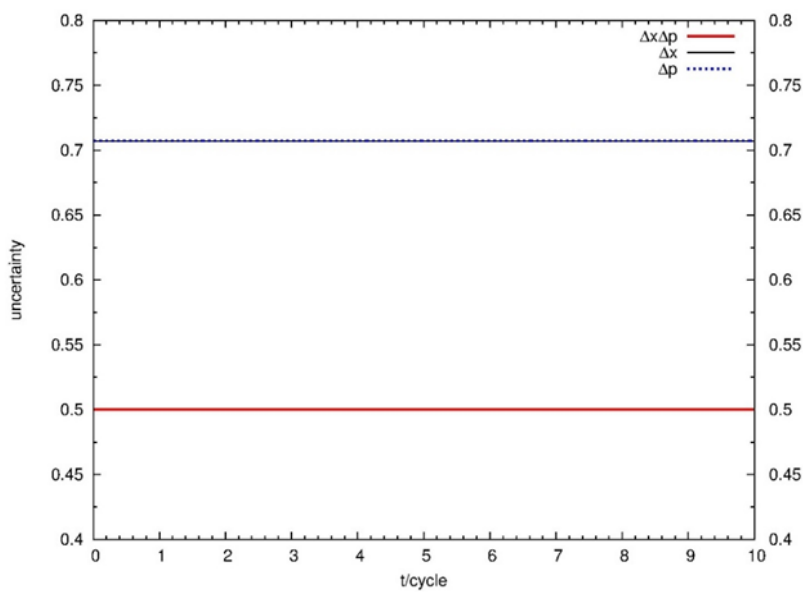

Fig. 1c

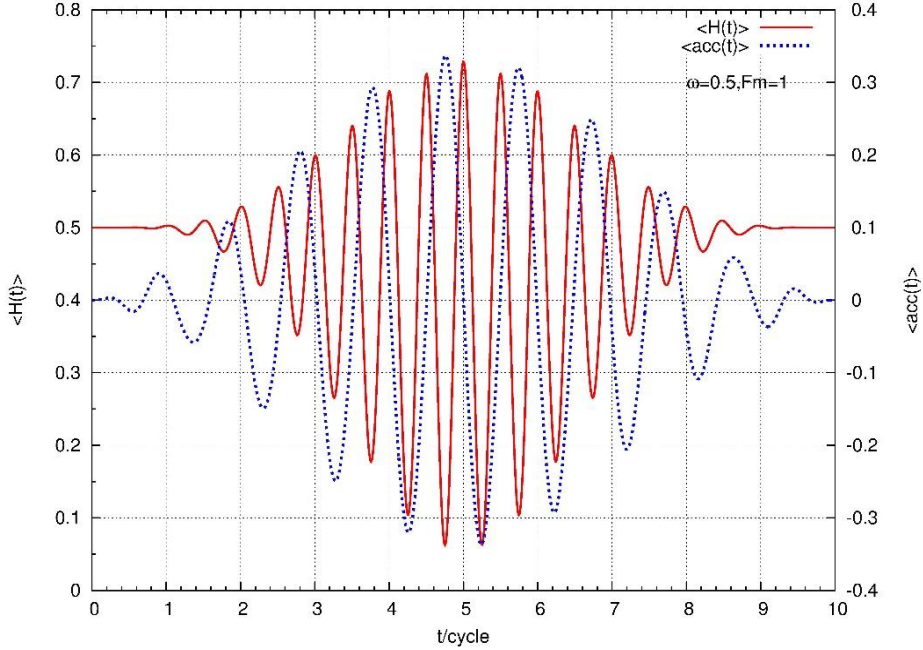

Next we show the results of resonant excitation. We use

$$\Omega = \omega, F_m = 1, T = 10 \text{ cycle} \qquad (38)$$

For stronger field, we need larger number of grids, yet the physics is the same. In Fig. 2a we show the corresponding results as Fig. 1a. We can see that the wave packet peak position again agrees with the classical d(t) faithfully. Interestingly that at the end of vanishing excitation, d(t) is not zero, instead, the dominant time behavior is $t \cos[\omega t]$ as shown above in analytic formula. The classical turning points are $x = \pm 1$ for the ground state energy. We can see that the peak position goes much far beyond the range due to the driving. In Fig. 2b we show the uncertainty and product. It shows the minimum uncertainty property through the excitation while the wave packet was energetically excited by the field F(t). Also we show in Fig.2c the time change of total energy and the acceleration function. Comparing with the nonresonant excitation, the two quantities keep increasing oscillatory to quite large values even after the end of pulse. Even the wave packet is driven far beyond the classical allowed region of the ground state, and with large acceleration and total energy, the wave packet still keeps the shape without distortion.

Fig.2a

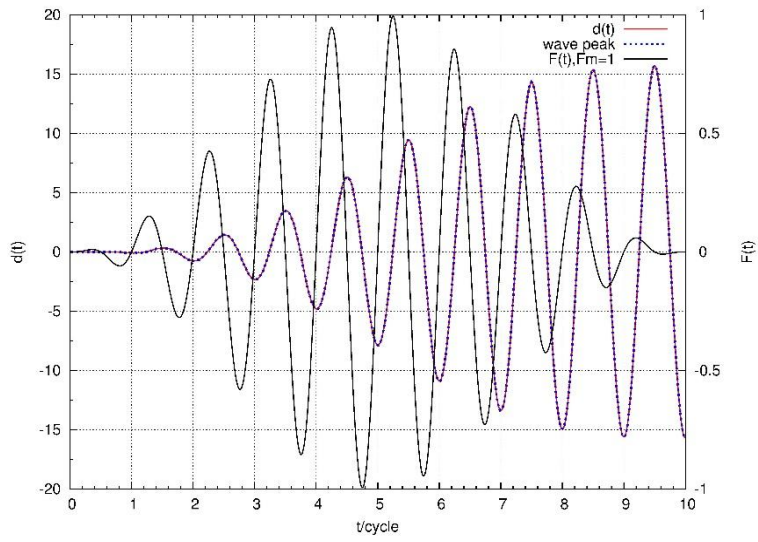

Fig. 2b

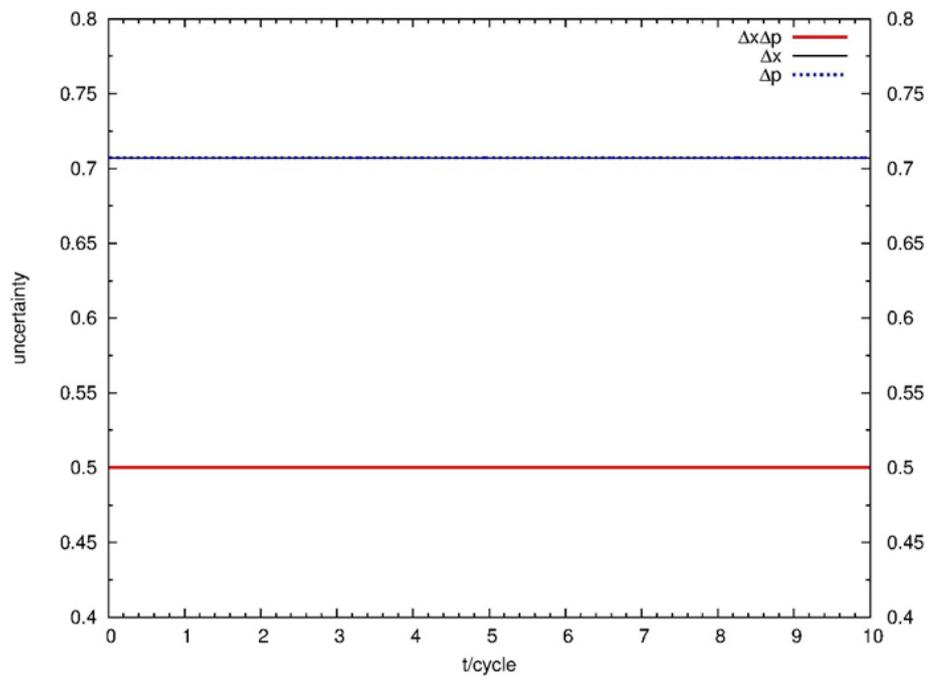

Fig. 2c

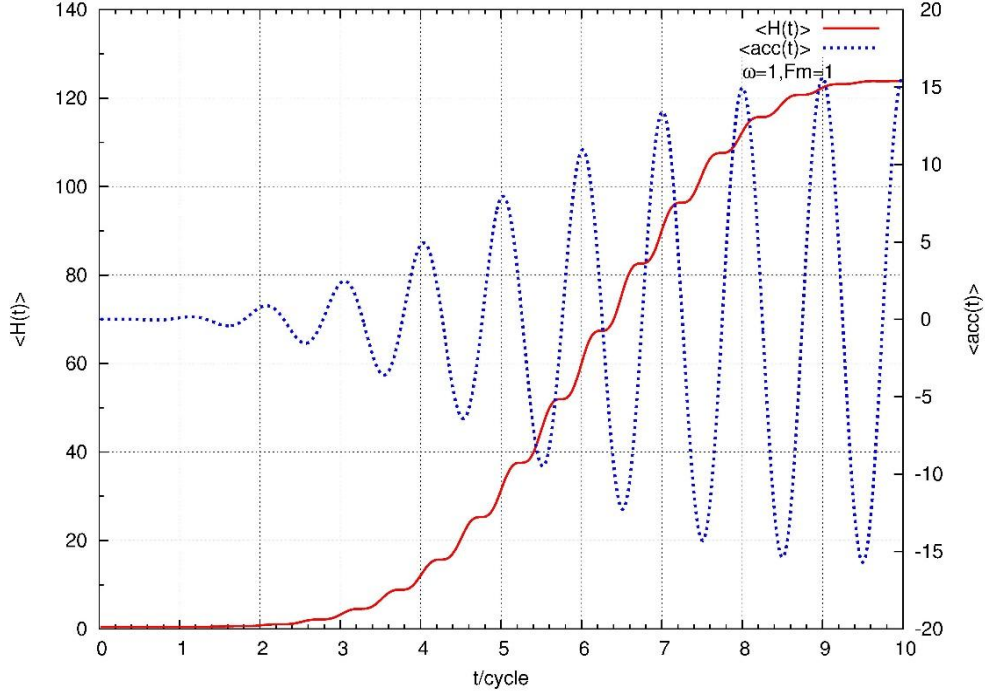

## 4. Conclusions

We have shown the decomposition method for analyzing the time evolution of quantum mechanical system. The eigenvalue equation of the wave function at each instant is used. That means the time development of a state is related to its static property at each instant. The eigenvalue equation, or the state-preserving operator $\widetilde{H}(t)$, in general includes the operator $\frac{p^2}{2m}$, hence the state-changing operator $H_c(t)$ is simpler. We can therefore exactly solve the driven harmonic oscillator. We also show that this system allows the existence of NSWPs. This is due to the state-changing operator $H_c(t)$ is linear in x and p.

We justify our theoretical results by numerical simulations. The peak position of wave packet follows the trajectory of classical driven oscillator. And the wave packet nonspreading property remains under resonant and nonresonant excitations.

This decomposition method can be applied to analyze the time evolution of general Hamiltonian systems as well.

# Acknowledgement

T.F. Jiang acknowledges the support of Taiwan Ministry of Science and Technology under the grant number MOST-103-2112-M-009-003.

Figure Captions

Fig.1a Calculated wave peak position in time (blue dots) and trajectory of driven classical oscillator d(t) (red line) of typical nonresonant excitation with pulse F(t) (solid line, label in the left axis). The parameters are described in context.

Fig.1b Time behavior of uncertainty and the uncertainty product.

Fig.1c Expectation value of time dependent Hamiltonian H(t) and the Ehrenfest's acceleration function.

Fig.2a   Calculated wave peak position in time (blue dots) and trajectory of driven classical oscillator d(t) (red line) of typical resonant excitation with pulse F(t) (solid line, label the in left axis). The parameters are described in context. Near the tail of the pulse, the field is vanishing but the wave peak position oscillates at $\sim \cos[\omega t]$ as predicted.

Fig.2b Time behavior of uncertainty and the uncertainty product of resonant excitation.

Fig.2c Expectation value of time dependent Hamiltonian H(t) and the Ehrenfest's acceleration function for the resonant excitation. Near the vanishing pulse tail, the wave packet is still accelerating.